\documentclass[pre,twocolumn,aps,showpacs]{revtex4-1}
\newcommand{\bec}[1]{\mbox{\boldmath $ #1$}}
\usepackage{graphicx}
\usepackage{bm}
\begin{document}
\title{Tangling clustering instability for small particles in temperature
stratified turbulence}
\author{T. Elperin$^1$}
\email{elperin@bgu.ac.il}
\homepage{http://www.bgu.ac.il/me/staff/tov}
\author{N. Kleeorin$^{1,2,3}$}
\email{nat@bgu.ac.il}
\author{M. Liberman$^{2,4}$}
\email{misha.liberman@gmail.com}
\homepage{http://michael-liberman.com/}
\author{I. Rogachevskii$^{1,2,3}$}
\email{gary@bgu.ac.il}
\homepage{http://www.bgu.ac.il/~gary}

\medskip
\affiliation{$^1$ The Pearlstone Center for
Aeronautical Engineering Studies, Department of
Mechanical Engineering, Ben-Gurion University of
the Negev, P. O. Box 653, Beer-Sheva
84105, Israel \\
 $^2$ Nordita, KTH Royal Institute of Technology
and Stockholm University, Roslagstullsbacken 23,
10691 Stockholm, Sweden \\
 $^3$ Department of Radio Physics, N.~I.~Lobachevsky State University of
Nizhny Novgorod, Russia \\
 $^4$ Moscow Institute of Physics and Technology,
Dolgoprudnyi, 141700, Russia}

\date{\today}
\begin{abstract}
We study tangling clustering instability of
inertial particles in a temperature stratified
turbulence with small finite correlation time. It
is shown that the tangling mechanism in the
temperature stratified turbulence strongly
increases the degree of compressibility of
particle velocity field. This results in the
strong decrease of the threshold for the
excitation of the tangling clustering instability
even for small particles. The tangling clustering
instability in the temperature stratified
turbulence is essentially different from the
inertial clustering instability that occurs in
non-stratified isotropic and homogeneous
turbulence. While the inertial clustering
instability is caused by the centrifugal effect
of the turbulent eddies, the mechanism of the
tangling clustering instability is related to the
temperature fluctuations generated by the
tangling of the mean temperature gradient by the
velocity fluctuations. Temperature fluctuations
produce pressure fluctuations and cause particle
accumulations in regions with increased
instantaneous pressure. It is shown that the
growth rate of the tangling clustering
instability is in $\sqrt{\rm Re} \, (\ell_0 /
L_T)^2 / (3 {\rm Ma})^4$ times larger than that
of the inertial clustering instability, where
${\rm Re}$ is the Reynolds number, ${\rm Ma}$ is
the Mach number, $\ell_0$ is the integral
turbulence scale and $L_T$ is the characteristic
scale of the mean temperature variations. It is
found that depending on the parameters of the
turbulence and the mean temperature gradient
there is a preferential particle size at which
the particle clustering due to the tangling
clustering instability is more effective. The
particle number density inside the cluster after
the saturation of this instability can be by
several orders of magnitude larger than the mean
particle number density. It is also demonstrated
that the evaporation of droplets drastically
change the tangling clustering instability, e.g.,
it increases the instability threshold in the
droplet radius. The tangling clustering
instability is of a great importance, e.g., in
atmospheric turbulence with temperature
inversions.
\end{abstract}


\maketitle

\section{Introduction}

Formation of spatial inhomogeneities in the
number density distribution of small inertial
particles in a turbulent flow (also called
particle clustering or preferential
concentration) has attracted considerable
attention in the past decades.
\cite{CST11,WA00,WA09,BE10,MBC12} The enhanced
number density of particles inside the cluster
may affect the particle interactions, their
dynamics and collisions. The dynamics of particle
collisions is relevant to many phenomena in
nature such as the raindrop formation and
atmospheric aerosols dynamics
\cite{S03,BH03,KPE07,SGG10}, as well as to
numerous industrial processes involving, e.g.,
sprays in diesel and jet engines \cite{PA02}.
Formation of clusters with enhanced number
density of particles may increase the rate of
particle collisions and coalescence.
\cite{S03,KPE07} This can significantly modify
the size and velocity distributions of the
droplets in the spray and affect the combustor
performance.

Clustering of inertial particles in a turbulent
flow is caused by the centrifugal effect, which
implies that the inertial particles are locally
accumulated in regions between the turbulent
eddies. These regions have a low vorticity, high
strain rate, and maximum fluid pressure.
Therefore, turbulent vortices act as small
centrifuges that push heavy particles to the
boundary regions between the eddies by the
inertial forces creating concentration
inhomogeneities. This effect is known as the
inertia-induced particle clustering \cite{M87}.
The inertial particle clustering in a turbulent
flow has been studied analytically
\cite{EKR96a,EKRS00,BF01,EKR02,MW05,EKR07,OV07,FH08,OL10},
numerically
\cite{BL03,CK04,CC05,CG06,YG07,BB07,AC08,BB10},
and experimentally
\cite{WA00,AC02,WH05,SA08,XB08,SSA08}.

In these study we distinguish between two types
of particle clustering. The first type is source
clustering related to the source term in the
equation for fluctuations of the particle number
density. Most of analytical studies of
preferential concentration are related to the
source inertial clustering. Another type of
particle clustering is associated with a
spontaneous breakdown of their homogeneous
spatial distribution due to the clustering
instability. \cite{EKR02,EKR07} The clustering
instability can be of great importance in
different practical applications involving
particle mixing and transport.

In the temperature stratified turbulence the
particle clustering is affected by turbulent
thermal diffusion. \cite{EKR96,EKR97} This
phenomenon causes accumulation of the inertial
particles in the vicinity of the mean temperature
minimum and results in the formation of
inhomogeneous mean particle number density
distributions. Turbulent thermal diffusion is a
purely collective phenomenon occurring in
temperature stratified turbulence and resulting
in the appearance of a non-zero mean effective
velocity of particles in the direction opposite
to the mean temperature gradient. A competition
between the turbulent thermal diffusion and
turbulent diffusion determines the conditions for
the formation of large-scale particle
concentrations in the vicinity of the mean
temperature minimum. The phenomenon of turbulent
thermal diffusion has been studied analytically
\cite{EKR96,EKR97,EKR00,EKR01,PM02,RE05,SSEKR09},
investigated by means of direct numerical
simulations \cite{HKRB12}, and detected in the
laboratory experiments in stably and unstably
stratified turbulent flows \cite{BEE04,EEKR06},
and also observed in atmospheric turbulence.
\cite{SSEKR09}

Particle clustering in the temperature stratified
turbulence can be much more effective than the
inertial particle clustering in isothermal
turbulence. \cite{EKR10} The reason for this is
that the mean temperature gradient in turbulent
flow is a strong source of the temperature
fluctuations which are correlated with the
fluctuations of fluid velocity and pressure. The
pressure fluctuations increase fluctuations of
the particles number density and enhance the rate
of formation of the particle clusters. Moreover,
tangling of the mean gradient of particle number
density (formed by the turbulent thermal
diffusion) generates additional fluctuations of
particle concentrations and contributes to the
particle clustering.

The steady-state regime of the tangling
clustering (i.e., the source tangling clustering)
in temperature stratified turbulence without
excitation of instability has been recently
studied experimentally and theoretically in
Ref.~\onlinecite{EKR10}. It was demonstrated that
in the laboratory stratified turbulence the
source tangling clustering is much more effective
than a pure inertial clustering that has been
observed in isothermal turbulence. In particular,
in the experiments in oscillating grid isothermal
turbulence in air without imposed mean
temperature gradient, the inertial clustering is
very weak for solid particles with the diameter
$\approx 10 \, \mu$m and Reynolds numbers based
on turbulent length scale and rms velocity, ${\rm
Re} =250$. In the experiments~\cite{EKR10} the
correlation function for the inertial clustering
in isothermal turbulence is significantly less
localized than that for the tangling clustering
in non-isothermal turbulence. The source tangling
clustering was studied in Ref.~\onlinecite{EKR10}
for inertial particles with small Stokes numbers
and with the material density that is much larger
than the fluid density.

The goal of the present paper is to investigate
theoretically another regime of the tangling
clustering, i.e., to study the tangling
clustering instability in the temperature
stratified turbulence. In this paper we show that
the tangling mechanism in the temperature
stratified turbulence strongly increases the
degree of compressibility of particle velocity
field and considerably enhances the growth rate
of the tangling clustering instability. For small
particles the tangling clustering instability may
result in the formation of small-scale particle
clusters with the number density of particles
exceeding the ambient average particle number
density by several orders of magnitude.

The paper is organized as follows. The
large-scale effects in particle transport in
temperature stratified turbulence are discussed
in Section II. The governing equations for
analysis of instability are given in Section III.
Solutions for the tangling clustering instability
without the source term in the equation for the
second moment of particle number density and with
the source term are analyzed in Sections IV and
V, respectively. The instability growth rate and
saturated value of the particle number density
inside a cluster are determined in Sections IV-V.
In Section VI we take into account an effect of
droplet evaporation on tangling clustering
instability. Finally, in Section VII we draw
conclusions and discuss the implications of the
tangling clustering instability.

\section{Particles in temperature stratified turbulence}

\subsection{Governing equations}

Advection-diffusion equation for the number
density $n_p(t,{\bm x})$ of inertial particles in
a turbulent flow reads \cite{CH43,AP81}:
\begin{eqnarray}
\frac{\partial n_p}{\partial t} + {\bm \nabla
\cdot} (n_p \,{\bm v}_p) = D_m\, \Delta n_p ,
 \label{B1}
\end{eqnarray}
where $D_m$ is the coefficient of molecular
(Brownian)  diffusion, ${\bm v}_p(t,{\bm x})$ is
the instantaneous particle velocity field. We use
a mean field approach in which the particle
number density and velocity, the fluid
temperature, density and pressure are decomposed
into the mean and fluctuating parts, where the
fluctuating parts have zero mean values.
Averaging Eq.~(\ref{B1}) over an ensemble of
turbulent velocity fields we obtain an equation
for the mean number density of particles:
\begin{eqnarray}
{\partial N \over \partial t} + \bec\nabla {\bf
\cdot} \, \left(N \, {\bm V}_p +\langle n \, {\bm
v}\rangle \right) = D_m \, \Delta N ,
 \label{BB1}
\end{eqnarray}
where $N=\langle n_p\rangle$ is the mean particle
number density, the angular brackets imply
ensemble averaging, $\langle n \, {\bm v}\rangle$
is the turbulent flux of particles, ${\bm
v}(t,{\bm x})$ are the fluctuations of the
particle velocity field and ${\bm V}_p$ is the
particle mean velocity. To obtain a closed
mean-field equation one needs to determine the
turbulent flux of particles. The equation for
fluctuations of the particle number density,
$n(t,{\bm x})= n_p(t,{\bm x}) - N(t,{\bm x})$,
then reads:
\begin{eqnarray}
\frac{\partial n}{\partial t} + {\bm \nabla
\cdot} (n \,{\bm v}- \langle n \,{\bm v}\rangle)
- D_m\, \Delta n = - ({\bm v} {\bm \cdot \nabla})
N - N \, {\bm \nabla \cdot} \,{\bm v} .
\nonumber\\
 \label{B3}
\end{eqnarray}

\subsection{Turbulent thermal diffusion}

Turbulent thermal diffusion results in formation
of a nonzero gradient of the mean particle number
density ${\bm \nabla} N$ in temperature
stratified turbulence. \cite{EKR96,EKR97} The
physical mechanism of turbulent thermal diffusion
is as follows. For the particles with the
material density $\rho_m \gg \rho$, their
velocity is determined by
\begin{eqnarray}
{d{\bm v} \over dt} = {{\bm u} - {\bm v} \over
\tau_s} + {\bm g},
 \label{D1}
\end{eqnarray}
where ${\bm u}$ is the fluid velocity field,
${\bm g}$ is the acceleration of gravity, $\tau_s
= m_p / 6 \pi \rho \, \nu a_p$ is the Stokes time
for the small spherical particles of the radius
$a_p$ and mass $m_p$, $\nu$ is the kinematic
viscosity, $\rho$ is the mean fluid density. For
small Stokes numbers, ${\rm St} =
\tau_s/\tau_\eta \ll 1$, solution of
Eq.~(\ref{D1}) has the following form (see, e.g.,
Ref.~\onlinecite{M87}):
\begin{eqnarray}
{\bm v} = {\bm u} - \tau_s\, \biggl[{\partial
{\bm u} \over \partial t} + ({\bm u} {\bm \cdot}
{\bm\nabla}) {\bm u} - {\bm g}\biggr] + {\rm
O}(\tau_s^2),
 \label{D2}
\end{eqnarray}
or introducing the dimensionless units, it can be
written in the dimensionless form:
\begin{eqnarray}
{\bm v} = {\bm u} - {\rm St}\, \biggl[{\partial
{\bm u} \over \partial t} + ({\bm u} {\bm \cdot}
{\bm\nabla}) {\bm u}  - {\tau_\eta^2{\bm g} \over
\ell_\eta}\biggr] + {\rm O}({\rm St}^2),
 \label{DD2}
\end{eqnarray}
where the distance is measured in the Kolmogorov
viscous scale units, $\ell_\eta=\ell_0/{\rm
Re}^{3/4}$, and the time is measured in the
Kolmogorov time scale units, $\tau_\eta=
\tau_0/{\rm Re}^{1/2}$. Here ${\rm Re}= \ell_0\,
u_0/\nu$ is the fluid Reynolds numbers, $u_0$ is
the characteristic turbulent velocity at the
integral scale $\ell_0$ of turbulent motions and
$\tau_0=\ell_0/u_0$. The terms in squared
brackets in Eq.~(\ref{D2}) describe the
difference between the local fluid velocity and
particle velocity arising due to the small but
finite inertia of the particle. For the turbulent
flow with low Mach numbers: ${\bm \nabla} {\bm
\cdot} \, {\bm u} \approx - \rho^{-1} \, ({\bm u}
{\bm \cdot} {\bm\nabla}) \rho \not= 0$. The
equation for ${\bm \nabla} {\bm \cdot} \, {\bm
v}$ can be easily obtained from the
equation~(\ref{D2}) and the Navier-Stokes
equation:
\begin{eqnarray}
{\bm \nabla} {\bm \cdot} \, {\bm v} &=& {\bm
\nabla} {\bm \cdot} \, {\bm u} - \tau_s \, {\bm
\nabla} {\bm \cdot} \,  \biggl( {d{\bm u} \over
dt} \biggr) + {\rm O}(\tau_s^2)
\nonumber\\
&=& - {1 \over \rho} \, ({\bm u} {\bm \cdot}
{\bm\nabla}) \rho + {\tau_s \over \rho}
\,{\bm\nabla}^2 p  + {\rm O}(\tau_s^2) ,
 \label{D3}
\end{eqnarray}
where $p$ are the fluid pressure fluctuations.
Due to inertia, particles inside the turbulent
eddies drift out to the boundary regions between
the eddies. These are regions with small velocity
fluctuations and maximum pressure fluctuations.
Consequently, particles are accumulated in the
regions with the maximum pressure fluctuations of
the turbulent fluid, i.e., ${\bm \nabla} {\bm
\cdot} \, {\bm v} \propto (\tau_s /\rho)
\,{\bm\nabla}^2 p \not=0$ even for the
incompressible fluid. For large Peclet numbers,
when the molecular diffusion of particles in
Eq.~(\ref{B1}) can be neglected, we can estimate
${\bm \nabla} {\bm \cdot} \, {\bm v} \propto - d
n_p / d t$. Therefore, inertial particles are
accumulated (i.e.,  $dn_p / dt \propto - (\tau_s
/\rho) \,{\bm\nabla}^2 p > 0)$ in regions with
maximum pressure of turbulent fluid, where
${\bm\nabla}^2 p < 0$. Similarly, there is an
outflow of inertial particles from the regions
with the minimum pressure of fluid. In case of
homogeneous and isotropic turbulence a drift from
regions with increased (decreased) concentration
of particles by a turbulent flow of fluid is
equiprobable in all directions.

On the contrary, in a temperature stratified
turbulence, the turbulent heat flux $\langle {\bm
u} \, \theta \rangle$ does not vanish. This
implies that the fluctuations of fluid
temperature, $\theta$, and velocity are
correlated, and, therefore, fluctuations of
pressure are correlated  with the fluctuations of
velocity due to a non-zero turbulent heat flux,
$\langle {\bm u} \, \theta \rangle \not = 0$. The
increased pressure of the surrounding fluid is
accompanied by the particles accumulation, and
the direction of the mean flux of particles
coincides with the direction of the heat flux
towards the minimum of the mean temperature.
\cite{EKR96,EKR10}

Equation for the mean number density  $N$ of
particles reads:
\begin{eqnarray}
{\partial N \over \partial t} + {\bm \nabla} {\bm
\cdot} \, \big[N \, ({\bm V}+{\bm W}_g) + {\bm
F}^{(n)} \big] =D_m \triangle N ,
 \label{D4}
\end{eqnarray}
where ${\bm V}_p={\bm V}+{\bm W}_g$ is the mean
particle velocity, ${\bm V}$ is the mean fluid
velocity, ${\bm W}_g=\tau_s {\bm g}$ is the
terminal fall velocity of particles, $D_m=k_B T/6
\pi \rho \nu a_p$ is the coefficient of molecular
(Brownian) diffusion, $k_B$ is the Boltzman
constant, $T$ and $\rho$ are the fluid mean
temperature and density, respectively. Hereafter
for simplicity we consider the case of a zero
mean fluid velocity, ${\bm V}=0$. The turbulent
flux of particles, ${\bm F}^{(n)} = \langle n \,
{\bm v} \rangle$, includes contributions of
turbulent thermal diffusion and turbulent
diffusion, i.e.,
\begin{eqnarray}
{\bm F}^{(n)} = {\bm V}^{\rm eff} \, N - D_{_{T}}
\, {\bm \nabla}  N .
 \label{D5}
\end{eqnarray}
Here $D_{_{T}} \approx \ell_0 \,u_0$ is the
coefficient of turbulent  diffusion, ${\bm
V}^{\rm eff}$ is the effective pumping velocity
caused by the turbulent thermal diffusion:
\begin{eqnarray}
{\bm V}^{\rm eff} = - \tau \, \langle {\bm v} \,
({\bm \nabla}  {\bm \cdot} \, {\bm v}) \rangle ,
\label{D6}
\end{eqnarray}
where $\tau$ is the turbulent correlation time.
Equation~(\ref{D6}) for the effective velocity
has been derived using different methods in
Refs.~\onlinecite{EKR96,EKR00,EKR01,PM02,RE05,SSEKR09}.
The expression~(\ref{D6}) can be obtained in a
simple way using the dimensional consideration.
Estimating the left hand side of Eq.~(\ref{B3})
as
\begin{eqnarray}
{\partial n \over\partial t} + {\bm \nabla} {\bm
\cdot} \, (n \,{\bm v}- \langle n \,{\bm
v}\rangle) - D_m {\bm\nabla}^2 n \approx {n
\over\tau},
 \label{D30}
\end{eqnarray}
we obtain an expression for the turbulent
component $n$ of the particle number density:
\begin{eqnarray}
n \approx - \tau \, {\bm \nabla} {\bm \cdot} \,
(N \, {\bm v}) = - \tau \, [N ({\bm \nabla} {\bm
\cdot} \, {\bm v}) + ({\bm v} {\bm \cdot}
{\bm\nabla}) N].
 \label{D31}
\end{eqnarray}
Therefore, the turbulent flux of particles
$F_i^{(n)} = \langle v_i \, n \rangle$ is given
by the following expression:
\begin{eqnarray}
F_i^{(n)} = - N \, \tau \, \langle v_i \,
({\bm \nabla} {\bm \cdot}  \, {\bm v}) \rangle  -
\tau \, \langle v_i v_j \rangle \nabla_j N \;,
\label{D7}
\end{eqnarray}
where the first term in the right hand side of
Eq.~(\ref{D7})  determines the turbulent flux of
particles due to the turbulent thermal diffusion:
$- N \, \tau \, \langle v_i \, ({\bm \nabla} {\bm
\cdot} \, {\bm v}) \rangle = V^{\rm eff}_i \, N$,
and the second term in the right hand side of
Eq.~(\ref{D7}) describes the contribution of
turbulent diffusion: $\tau \, \langle v_i v_j
\rangle \nabla_j N = D_{_{T}} \,\nabla_i N$.

A detailed analysis \cite{EKR96,EKR10}, using
equation of state for an ideal gas with adiabatic
index $c_{\rm p}/c_{\rm v}$ (the ratio of
specific heats) and applying the identity $\tau_s
= \rho \, W_g \, L_P / P$, yields the effective
velocity in the following form:
\begin{eqnarray}
{\bm V}^{\rm eff} = - \alpha \, D_{_{T}} \,
{{\bm \nabla} T \over T},
 \label{D8}
\end{eqnarray}
where $\alpha$ is
\begin{eqnarray}
\alpha = 1 + {{\rm St} \, \ln({\rm Re}) \over
\sqrt{\rm Re} \, \, {\rm Ma}^2}.
 \label{D9}
\end{eqnarray}
Here $L_P^{-1} = |{\bm \nabla} P|/P$, ${\rm
Ma}=u_0/c_s$ is the Mach number and $c_s$ is the
sound speed. For gases and non-inertial particles
$\alpha =1$. A steady-state solution of
Eq.~(\ref{D4}) is given by the following formula:
\begin{eqnarray}
{N(z) \over N_0} = \left[{T(z)\over
T_0}\right]^{-{\alpha D_T \over D_m+D_T}} \, \exp
\left[-\int_{z_0}^z \, {W_{\rm g} \over D_m+D_T}
\,\,d z' \right] ,
\nonumber\\
 \label{D10}
\end{eqnarray}
where $N_0=N(z=z_0)$ and $T_0=T(z=z_0)$ are the
mean number density of particles and the mean
fluid temperature, respectively, calculated at
the boundary $z=z_0$. If there is a gradient of
temperature in a vertical $z$ direction,
Eq.~(\ref{D10}) implies that small particles are
accumulated in the vicinity of the mean
temperature minimum. This causes formation of
large-scale inhomogeneous distributions of the
mean particle number density.

\section{Governing equations for analysis of instability}

Let us study fluctuations of the particle number
density. The methodology and approach used for
investigation of the tangling clustering
instability are similar to the methodology and
approach used for study of the inertial
clustering instability. \cite{EKR02,EKR07} We
apply the path-integral approach for random
compressible flow with the small yet finite
correlation time for the derivation of the
equation for the correlation function of the
particle number density. This approach is
described comprehensively in
Refs.~\onlinecite{EKR00,EKR01}. The equation for
the two-point second-order correlation function
of the particle number density,
\begin{eqnarray*}
\Phi(t,{\bm R})=\langle n(t,{\bm x}) n(t,{\bm
x}~+~{\bm R}) \rangle,
\end{eqnarray*}
is given by
\begin{eqnarray}
{\partial \Phi \over \partial t} &=& \big[B({\bm
R}) + 2 {\bm U}^{(A)}({\bm R})\cdot {\bm\nabla} +
\hat D_{ij}({\bm R}) \nabla_{i} \nabla_{j}\big]
\, \Phi(t,{\bm R})
\nonumber\\
&& + I({\bm R}) ,
 \label{B2}
\end{eqnarray}
where ${\bm U}^{(A)}({\bm R}) = (1/2) \,
\big[{\bm U}({\bm R}) - {\bm U}(-{\bm R})\big]$,
\begin{eqnarray}
\hat D_{ij}&=& 2 D_m \delta_{ij} +
D_{ij}^{^{T}}(0) - D_{ij}^{^{T}}({\bm R}) ,
 \label{B4}\\
D_{ij}^{^{T}}({\bm R}) &\approx& 2
\int_{0}^{\infty} \langle v_{i}
\big[0,{\bm\xi}(t,{\bm x}|0)\big] \,
v_{j}\big[\tau,{\bm\xi}(t,{\bm x}+{\bm
R}|\tau)\big] \rangle \,d \tau ,
\nonumber\\
 \label{B5}\\
B({\bm R}) &\approx& 2 \int_{0}^{\infty} \langle
b\big[0,{\bm\xi}(t,{\bm x}|0)\big]
\,b\big[\tau,{\bm\xi}(t,{\bm x}+{\bm
R}|\tau)\big] \rangle \,d \tau ,
\nonumber\\
 \label{B6}\\
U_{i}({\bm R}) &\approx& -2 \int_{0}^{\infty}
\langle v_{i}\big[0,{\bm\xi}(t,{\bm x}|0) \big]
\,b\big[\tau,{\bm\xi}(t,{\bm x}+{\bm
R}|\tau)\big] \rangle \,d \tau
\nonumber\\
 \label{B7}
\end{eqnarray}
(see for details of derivations
Ref.~\onlinecite{EKR02}). Here $b={\rm div} \,
{\bm v}$, $\, D_{ij}^{^{T}}({\bm R})$ is the
scale-dependent turbulent diffusion tensor,
$\delta_{ij}$ is the Kronecker tensor, $I({\bm
R})$ is the source of particle number density
fluctuations and $\langle ... \rangle$ denotes
averaging over the statistics of turbulent
velocity field and the Wiener process ${\bm
w}(t)$. The Wiener trajectory ${\bm\xi}(t,{\bm
x}|s)$ in the expressions for the turbulent
diffusion tensor $D_{ij}^{^{T}} ({\bm R})$ and
other transport coefficients is defined as
follows:
\begin{eqnarray}
{\bm\xi}(t,{\bm x}|s) &=&{\bm x} - \int^{t}_{s}
{\bm v} [\tau,{\bm\xi}(t,{\bm x}|\tau)] \,\,d
\tau - \sqrt{2 D_m} \, {\bm w}(t-s) ,
\nonumber\\
 \label{B8}
\end{eqnarray}
where ${\bm w}(t)$ is the Wiener random process
which describes the Brownian motion (molecular
diffusion) and has the following properties:
$\langle {\bm w}(t) \rangle_{\bm w}=0\,, $ $\,
\langle w_i(t+\tau) w_j(t) \rangle_{\bm w}= \tau
\delta _{ij}$, and $ \langle \dots \rangle_{\bm
w} $ denotes the mathematical expectation over
the statistics of the Wiener process. The
velocity $v_i[\tau, {\bm\xi}(t,{\bm x}|\tau)]$
describes the Eulerian velocity calculated at the
Wiener trajectory.

To simplify the averaging procedure in derivation
of Eq.~(\ref{B2}) we used a model of random
velocity field which fully looses memory at
random instants. The velocity fields before and
after renewal are assumed to be statistically
independent. Between the renewals the velocity
field can be random with its intrinsic
statistics. To obtain a statistically stationary
random velocity field we assumed that the
velocity fields between renewals have the same
statistics. The random renewal instants destroy
stationarity of the velocity field. On the other
hand, between the random renewal instants the
velocity field is stationary in statistical
sense. To perform calculations in the closed form
we assumed that the random renewal times are
determined by a Poisson process. We also
considered a model of a random velocity field
where Lagrangian trajectories, i.e., the
integrals $ \int {\bm v}(\mu, {\bm \xi}) \,d \mu$
and $\int b(\mu,{\bm \xi}) \,d \mu$ have Gaussian
statistics.

This model employs three random processes: (i)
the Wiener random process which describes
Brownian motions, i.e., the molecular diffusion;
(ii) Poisson process for random renewal times;
(iii) the random velocity field between the
renewals. This model reproduces important
features of some real turbulent flows. For
example, the interstellar turbulence which is
driven by supernovae explosions, loses memory in
the instants of explosions (see, e.g.,
Ref.~\onlinecite{LST00}). Such flows also can be
reproduced in direct numerical simulations.

Equation~(\ref{B2}) with $I({\bm R})=0$ and for a
delta-correlated in time random incompressible
$(b=0)$ velocity field was derived by Kraichnan.
\cite{K68} In this case: $B({\bm R})=\nabla_{i}
\nabla_{j} \hat D_{ij}({\bm R})$ and
$U_i^{(A)}({\bm R})=\nabla_{j} \hat D_{ij}({\bm
R})$, and Eq.~(\ref{B2}) is reduced to $\partial
\Phi /\partial t= \nabla_{i} \nabla_{j} [\hat
D_{ij}({\bm R}) \Phi(t,{\bm R})]$. For a
turbulent compressible flow with a finite
correlation time Eq.~(\ref{B2}) was derived using
a stochastic calculus. \cite{EKR02} In
particular, Wiener path integral representation
of the solution of the Cauchy problem for
Eq.~(\ref{B1}), the Feynman-Kac formula and
Cameron-Martin-Girsanov theorem were used for the
derivation of Eq.~(\ref{B2}).
\cite{EKR00,EKR01,ZRS90}

The source function $I({\bm R})$ in
Eq.~(\ref{B2}) is related to the two source terms
$- ({\bm v} {\bm \cdot \nabla}) N - N \, {\bm
\nabla \cdot} \,{\bm v}$ in the right hand side
of Eq.~(\ref{B3}), and the explicit expression
for $I({\bm R})$ is as follows (see
Ref.~\onlinecite{EKR10}):
\begin{eqnarray}
I({\bm R}) &=& B({\bm R}) N^2 + {\bm
U}^{(S)}({\bm R}) \cdot {\bm\nabla} N^2
\nonumber\\
&& + {3 \over 4} D_{ij}^{^{T}}({\bm R}) \,
(\nabla_{i} N) \, (\nabla_{j}N) ,
 \label{B9}
\end{eqnarray}
where ${\bm U}^{(S)}({\bm R}) = (1/2) \,
\big[{\bm U}({\bm R}) + {\bm U}(-{\bm R})\big]$
and $\nabla_i^{({\bm x})} \nabla_j^{({\bm y})}
N(t,{\bm x}) N(t,{\bm y}) = (3/4) \, (\nabla_{i}
N) \, (\nabla_{j}N)$, and $\nabla_i \equiv
\nabla_i^{({\bm R})}$.

The meaning of the turbulent transport
coefficients  $B({\bm R})$ and ${\bm U}({\bm R})$
is as follows. The function $B({\bm R})$ is
determined by the compressibility of the particle
velocity field. The vector ${\bm U}({\bm R})$
determines a scale-dependent drift velocity which
describes transport of fluctuations of particle
number density from smaller scales to larger
scales, i.e., in the regions with larger
turbulent diffusion. The scale-dependent tensor
of turbulent diffusion $D_{ij}^{^{T}}({\bm R})$
is equal to the tensor of the molecular
(Brownian) diffusion in very small scales, while
in the vicinity of the maximum scale of turbulent
motions it coincides with the tensor of turbulent
diffusion. It should be noticed also, that if
${\bm \nabla} N \not= 0$ a nonzero source term
$I({\bm R})$ causes the production of the
particle number density fluctuations due to the
tangling of the mean particle number density by
the turbulent velocity field.

\subsection{Degree of compressibility}

If the turbulent velocity field is not
delta-correlated  in time (e.g., the correlation
time is small yet finite), the tensor of
turbulent diffusion, $D_{ij}^{^{T}} ({\bm R})$,
is compressible, i.e., $(\partial / \partial R_i)
D_{ij}^{^{T}} ({\bm R}) \not= 0$. The parameter
$\sigma_{_{T}}$ that characterizes the degree of
compressibility of the tensor of turbulent
diffusion, is defined as follows:
\begin{eqnarray}
\sigma_{_{T}} \equiv \frac{\nabla_i \nabla_j
D^{^{\rm T}}_{ij}({\bm R})} {\nabla_i\nabla_j
D^{^{\rm T}} _{mn}({\bm R})
\epsilon_{imp}\epsilon_{jnp} } \approx {\langle
({\bm\nabla} {\bm \cdot} \tilde{\bm \xi})^{2}
\rangle \over \langle ({\bm\nabla} {\bm \times}
\tilde{\bm \xi})^{2} \rangle} \;,
 \label{B10}
\end{eqnarray}
where $\epsilon_{ijk}$ is the fully antisymmetric
Levi-Civita unit tensor, $\tilde{\bm \xi}
={\bm\xi} - {\bm x}$ with $|t-s| \gg \tau_0$. If
the turbulent velocity field is a
delta-correlated in time random  process, then
${\bm\nabla} {\bm \cdot} \tilde{\bm \xi} = -
({\bm\nabla} {\bm \cdot} {\bm v}) \, (t-s)$, and
${\bm\nabla} {\bm \times} \tilde{\bm \xi} = -
({\bm\nabla} {\bm \times} {\bm v}) \, (t-s)$,
and, hence, the expression for $\sigma_{_{T}}$
reads:
\begin{eqnarray}
\sigma_{_{T}} = {\langle ({\bm\nabla} {\bm \cdot}
{\bm v})^{2} \rangle  \over \langle ({\bm\nabla}
{\bm \times} {\bm v})^{2} \rangle} \equiv
\sigma_{\rm v} ,
 \label{B11}
 \end{eqnarray}
where $\sigma_{\rm v}$ is the degree of
compressibility of the particle velocity field.
The parameter $\sigma_{\rm v}$ depends on the
Stokes number \cite{EKR07}:
\begin{eqnarray}
\sigma_{\rm v} = {(8/3) {\rm St}^2  \over 1 +
\beta {\rm St}^2} ,
 \label{B11}
 \end{eqnarray}
where the coefficient $\beta \sim 1$.

For small finite correlation time of turbulent
velocity field (i.e., for small Strouhal numbers,
${\rm Sr} = \tau_c \sqrt{\langle {\bm u}^{2}
\rangle}/ \ell \ll 1$, the parameter
$\sigma_{_{T}}$ can be estimated using Eq.~(C12)
in Ref.~\onlinecite{EKR02} (see also
Ref.~\onlinecite{EKR10}) as follows:
\begin{eqnarray}
\sigma_{_{T}} = \sigma_{\rm v} + {2 \, {\rm
Sr}^2\over 3} \,  \Big(1+ {913 \, \sigma_{\rm
v}^2 \over 12 \, (1 +\sigma_{\rm v})} \Big) +
O({\rm Sr}^4) \;,
 \label{B12}
 \end{eqnarray}
where $\tau_c$ is the correlation time of random
velocity field. Here the condition Sr $\ll 1$ is
supposed to be valid in the whole inertial range
of scale.

The mechanism of coupling related to the tangling
of the gradient of the mean temperature gradient
is quite robust. The properties of the tangling
are not very sensitive to the exponent of the
energy spectrum of the background turbulence.
Anisotropy effects do not introduce new physics
in the clustering process because the main
contribution to the tangling clustering
instability is at the Kolmogorov (viscous) scale
of turbulent motions, where turbulence can be
considered as nearly isotropic, while anisotropy
effects can be essential in the vicinity of the
maximum scales of the turbulent motions. Using
these arguments, we consider the tensor $D^{^{\rm
T}}_{ij} ({\bm R})$ for isotropic and homogeneous
turbulent flow in the following form:
\begin{eqnarray}
D^{^{\rm T}}_{ij} ({\bm R}) &=& D_{_{\rm T}} \,
\biggl[ [F(R) + F_c(R)] \delta_{ij} + R F'_c \,
{R_i R_j \over R^2}
\nonumber\\
 && + {R F' \over 2} \left(\delta_{ij} - {R_i R_j
\over R^2} \right) \biggr] ,
 \label{B20}
\end{eqnarray}
where $F(0) = 1 - F_c(0)$ and  $F'=dF/dR$. The
function $F_c(R)$ describes the compressible
(potential) component, whereas $F(R)$ corresponds
to vortical (incompressible) part of the
turbulent diffusion tensor.

\subsection{Derivation of expression for the function $B({\bm R})$}

Taking into account the equation of state of an
ideal gas we obtain $p/P = \varrho/\rho +
\theta/T + O(\varrho \theta/\rho T)$. For small
Stokes numbers, ${\bm \nabla} {\bm \cdot} \, {\bm
v} \approx (\tau_s/\rho) \, {\bm\nabla}^2 p$.
This allows us to estimate $B({\bm R})$ as
\begin{eqnarray}
B({\bm R}) &\approx& {2 \tau_s^2 \over \rho^2} \,
\langle  \tau \big[{\bm\nabla}^2 p({\bm x})
\big]\, \big[{\bm\nabla}^2 p({\bm y})\big]
\rangle
\nonumber\\
&\approx&  {2 \tau_s^2 \over \rho^2} \, {P^2
\over T^2} \, \langle \tau \big[{\bm\nabla}^2
\theta({\bm x})\big] \, \big[{\bm\nabla}^2
\theta({\bm y})\big] \rangle,
 \label{R1}
\end{eqnarray}
where $\rho, T, P$ and $\varrho, \theta, p$ are
the mean and fluctuations of the fluid density,
temperature, and pressure, respectively, and
${\bm\nabla}^2 p({\bm x}) = \big[{\bm
\nabla}^{({\bm x})}  \big]^2 p({\bm x})$.
Hereafter we omit the argument $t$ in the
correlation function. In ${\bm k}$ space the
correlation function $\langle \tau
\big[{\bm\nabla}^2 \theta({\bm x})\big] \,
\big[{\bm\nabla}^2 \theta({\bm y})\big] \rangle$
reads:
\begin{eqnarray}
\langle \tau \big[{\bm\nabla}^2 \theta({\bm
x})\big] \,  \big[{\bm\nabla}^2 \theta({\bm
y})\big] \rangle &=& \int \tilde \tau(k) \, k^4
\, \langle \theta({\bm k}) \, \theta(-{\bm k})
\rangle
\nonumber\\
&& \times \exp \big(i {\bm k} {\bm \cdot} {\bm
R}\big) \, d{\bm k} .
 \label{R2}
\end{eqnarray}
Taking into account that $\langle \theta({\bm k})
\, \theta(-{\bm k}) \rangle = \langle \theta^2
\rangle \tilde E_\theta(k) / 4 \pi k^2$, and
integrating in ${\bm k}$ space we arrive to the
following expressions for the functions $B({\bm
R})$:
\begin{eqnarray}
B({\bm R}) \approx {2 \, {\rm St}^2 \, c_s^4 \,
\over 3 \, \tau_\eta \, u_0^4} \,\left({\ell_0
{\bm \nabla} T \over T} \right)^2 \, {\rm
Re}^{1/2} ,
 \label{R3}
\end{eqnarray}
where $\tilde E_\theta(k)=(2/3) \, k_0^{-1} \,
(k/k_0)^{-5/3}$ is the spectrum function of the
temperature fluctuations for $k_0 \leq k \leq
\ell_\eta^{-1}$, with $k_0=\ell_0^{-1}$ and
$\tilde \tau(k)=2 \tau_0 \, (k/k_0)^{-2/3}$. To
determine $\langle\theta^2 \rangle$ we used the
budget equation for the temperature fluctuations
$E_\theta=\langle \theta^2 \rangle/2$:
\begin{eqnarray}
{D E_\theta \over Dt} + {\rm div} \, {\bf
\Phi}_\theta = - ({\bf F} {\bf \cdot}
\bec{\nabla}) T - \varepsilon_\theta ,
 \label{R4}
\end{eqnarray}
which for homogeneous turbulence in a steady
state yields:
\begin{eqnarray}
\langle \theta^2 \rangle=- 2 \, \tau_0 \,({\bf F}
{\bf \cdot} {\bm \nabla}) T = {2 \over 3}
\left(\ell_0 {\bm \nabla} T\right)^2,
 \label{R5}
\end{eqnarray}
where $F_i = \langle u_i \theta \rangle = -
D_{_{T}}^{(\theta)} \nabla_i T$ is the turbulent
heat flux, $D_{_{T}}^{(\theta)}=u_0 \ell_0 /3$ is
the coefficient of the turbulent diffusion of the
temperature fluctuations and the dissipation rate
of $E_\theta$ is $\varepsilon_\theta = \langle
\theta^2 \rangle /2 \tau_0$ (see, e.g.,
Ref.~\onlinecite{ZEKR07}).

On the other hand, the function $B({\bm R})$ at
very small scales has a universal form:
\begin{eqnarray}
B({\bm R}) = {20 \, \sigma_{\rm v} \over
\tau_\eta \, (1 + \sigma_{\rm v})} \approx {20 \,
\sigma_{\rm v} \over \tau_\eta} \approx {160 \,
{\rm St}_{\rm eff}^2 \over 3\tau_\eta},
 \label{R6}
\end{eqnarray}
because at this scales the velocity field is
smooth and nearly isotropic. Here we introduced
the effective Stokes number using Eqs.~(\ref{R3})
and~(\ref{R5}):
\begin{eqnarray}
&& {\rm St}_{\rm eff} ={\rm St} \, \Gamma,
\nonumber\\
&& \Gamma({\rm Ma}, {\rm Re}, \ell_0/L_T) =
\left[1 + {{\rm Re}^{1/2} \over 81 \, {\rm
Ma}^{4}} \,\left({\ell_0 {\bm \nabla} T \over T}
\right)^2\right]^{1/2} ,
\nonumber\\
 \label{R9}
\end{eqnarray}
where $L_T$ is the characteristic scale of the
mean temperature variations. The case of
$\Gamma=1$ corresponds to the inertial clustering
instability. To derive Eq.~(\ref{R6}) we took
into account that for a Gaussian velocity field
\cite{EKR02}:
\begin{eqnarray}
\langle ({\bm\nabla} {\bm \cdot} {\bm v})^{2}
\rangle  = {80 \over 3 \,\tau_\eta^2} \, {\rm
St}_{\rm eff}^2 ,
 \quad
\langle ({\bm\nabla} {\bm\times} {\bm v})^{2}
\rangle = {10 \over \tau_\eta^2}.
 \label{R7}
 \end{eqnarray}
Equation~(\ref{R7}) yields:
\begin{eqnarray}
\sigma_{\rm v} \equiv {\langle ({\bm\nabla} {\bm
\cdot} {\bm v})^{2} \rangle \over \langle
({\bm\nabla} {\bm \times} {\bm v})^{2} \rangle}
={8 \over 3} \, {\rm St}_{\rm eff}^2 \ll 1 .
 \label{R8}
\end{eqnarray}
Taking typical parameters for atmospheric
turbulence ${\rm Re}=10^7$, $u_0=1$ m/s and
$\ell_0=100$ m within the temperature inversion,
such that the mean temperature gradient, $|{\bm
\nabla} T|$, is of the order of 1 K per 100 m, we
obtain  $\Gamma \approx 2.5 \times 10^3$, or for
${\rm Re}=10^6$, $u_0=0.3$ m/s and $\ell_0=30$ m
the parameter $\Gamma$ is $\Gamma \approx 5
\times 10^3$.

\section{Solution for tangling clustering instability
with zero source term $(I=0)$}

It is convenient to rewrite Eq.~(\ref{B2}) for
the two-point second-order correlation function
$\Phi(R)$ in a non-dimensional form  with
coordinate $R$ measured in units of the
Kolmogorov scale, $\ell_\eta$, time in units of
the Kolmogorov time, $\tau_\eta$ and the function
$\Phi$ in units of $N^2$:
\begin{eqnarray}
{\partial \Phi\over \partial t} &=& {1 \over
M(R)} \left[\Phi'' + 2 \, \left({1 \over R} +
\chi(R) \right) \, \Phi' \right] + 2 \tilde U R
\, \Phi'
\nonumber\\
&& + B(R) \, \Phi + I(R) ,
 \label{T1}
\end{eqnarray}
where
\begin{eqnarray}
&&{1 \over M(R)} = {2 \over {\rm Sc}} + {2 \over
3} [1 - F - (R F_c)']\;,
\nonumber\\
&& \chi(R) = - {M(R) \over 3} (F - 2 F_c)' \;,
 \label{B15}\\
&& I(R) = B(R) + {4 (\alpha-1)^2 \over 3} \,
\,\left({\ell_0 {\bm \nabla} T \over T} \right)^2
\nonumber\\
&& \quad \quad \times \left(1 - {3 \over 2 M(R)}
- {R \, \chi(R) \over M(R)}\right) ,
 \label{B16}
\end{eqnarray}
${\rm Sc} = \nu / D_m$ is the Schmidt number and
${\bm U} =  \tilde U(R) \, {\bm R}$. Typically,
in many applications, e.g., in the atmospheric
turbulence, ${\rm Sc} \gg 1$ for small inertial
particles. The two-point correlation function
$\Phi(R)$ satisfies the following boundary
conditions: $\Phi'(R=0) = 0$ and $\Phi(R \to
\infty) = 0$. This function has a global maximum
at $R=0$ and therefore it satisfies the
conditions:
\begin{eqnarray*}
\Phi''(R=0) < 0\,, \quad  \Phi(R=0) >  |\Phi (R>0)| \; .
\end{eqnarray*}

For a steady-state regime and in the absence of
the tangling clustering instability the solution
of Eq.~(\ref{T1}) was obtained in
Ref.~\onlinecite{EKR10} for $I(R) \not=0$. In
this section we will consider solution of
Eq.~(\ref{T1}) for the case of the tangling
clustering instability and $I(R)=0$. Since the
Schmidt number, ${\rm Sc}=\nu/D_m \gg 1$, the
molecular diffusion scale is much less than the
viscous Kolmogorov scale. A general form of the
turbulent diffusion tensor in the viscous range
of scales is obtained taking into account that
$F(R) = (1 - R^2) / (1 + \sigma_{_{T}})$ and
$F_c(R) = \sigma_{_{T}} (1 - R^2) / (1 +
\sigma_{_{T}})$, which yields:
\begin{eqnarray}
&& D^{^{\rm T}}_{ij}({\bm R})=  C_{1} R^{2}
\delta_{ij} + C_{2} R_i R_j \,,
\label{C1}\\
&&C_{1} = {2 (2 + \sigma_{_{T}}) \over 3(1 +
\sigma_{_{T}})}\,, \quad  C_{2} = {2
(2\sigma_{_{T}} - 1) \over 3(1 +
\sigma_{_{T}})}\; ,
 \label{C2}
\end{eqnarray}
and the other functions in this range of scales
are $\tilde U = 20 \, \sigma_{\rm v}/3(1 +
\sigma_{\rm v})$ and $B = 20 \, \sigma_{\rm v}/(1
+ \sigma_{\rm v})$.

In the {\it molecular diffusion range of scales},
$a_p / \ell_\eta \leq R \leq 1 / \sqrt{{\rm
Sc}}$, all terms $\propto R^2$ are small and can
be neglected. Note that we consider the case when
the particle radius, $a_p$, is the minimum scale
of the problem, so that  $a_p \leq \ell_D$, where
$\ell_D=\ell_0 / {\rm
Pe}^{3/4}=\ell_\eta/\sqrt{{\rm Sc}}$ is the
molecular diffusion scale, and ${\rm Pe}=u_0
\ell_0/D_m$ is the Peclet number. The solution of
Eq.~(\ref{T1}) in this range reads:
\begin{eqnarray}
\Phi(R) = \left(1 - {{\rm Sc} (B - \gamma
\tau_\eta) \over 12} R^{2}\right) \exp(\gamma t),
 \label{C3}
\end{eqnarray}
where $B > \gamma \tau_\eta$ and $\gamma$ is the
growth rate of the tangling clustering
instability.

In the {\it turbulent diffusion region of
scales}, $1 / \sqrt{{\rm Sc}} \ll R \ll 1$, the
molecular diffusion term $\propto 1 / {\rm Sc}$
is negligible, and we seek for the solution of
Eq.~(\ref{T1}) in this region in the following
form:
\begin{eqnarray}
\Phi(R) \propto R^{-\beta} \exp(\gamma t).
 \label{C4}
\end{eqnarray}
Using the Corrsin integral, $\int_{0}^{\infty}
R^{2} \Phi(R) \,d R = 0$, we obtain that $\beta =
\lambda \pm i \kappa$ is a complex number,
$\kappa^2 > 0$, where
\begin{eqnarray}
&& \kappa^2 = {4 (B-\gamma\tau_\eta)(C_1+C_2) -
(C_{1} - C_{2} + 2 \tilde U)^{2} \over 4 (C_{1} +
C_{2})^2},
 \label{C5}\\
&& \lambda = {C_{1} - C_{2} + 2 \tilde U  \over 2
(C_{1} + C_{2})} = {3 - \sigma_{_{T}} \over 2(1 +
3\sigma_{_{T}})}+ {10 \sigma_{\rm v} \over 1 +
\sigma_{\rm v}} \left({1 + \sigma_{_{T}} \over
1+3\sigma_{_{T}}}\right) .
\nonumber\\
 \label{CC31}
\end{eqnarray}
Hence the real part of solution~(\ref{C4}) is
reduced to
\begin{eqnarray}
\Phi(R) = C R^{-\lambda} \cos \left(\kappa \ln R
+ \varphi\right) \exp(\gamma t) ,
 \label{C31}
\end{eqnarray}
where $C$ is the constant.

Since the correlation function $\Phi(R)$ has a
global maximum in $R = a_p/\ell_\eta \ll 1$, the
parameter $\sigma_{T} \leq 3$. The function
$\Phi(R)$ sharply decreases with the increase of
$R$, for $R \gg 1$. The growth rate of the second
moment of particles number density and the
constant $C$ can be obtained by matching the
correlation function $\Phi(R)$ and its first
derivative $\Phi'(R)$ at the boundaries of the
above ranges, i.e., in the points $R = 1 /
\sqrt{{\rm Sc}}$ and $R = 1$. The matching yields
$\kappa / 2(C_{1} + C _{2}) \approx \pi \, m /
\ln {\rm Sc}$ (where $m=2k+1$), and the growth
rate for the m-th mode of the tangling clustering
instability is given by the following formula
\begin{eqnarray}
\gamma_m &=& {1 \over 3 (1 + 3 \sigma_{_{T}})}
\biggl[{200 \sigma_{\rm v} (\sigma_{_{T}} -
\sigma_{\rm v}) \over (1 + \sigma_{\rm v})^2} -
{(3 - \sigma_{_{T}})^{2} \over 2 (1 +
\sigma_{_{T}})}
\nonumber\\
&&- {2 \pi^2 m^2 (1 + 3 \sigma_{_{T}})^{2} \over
(1 + \sigma_{_{T}}) \ln^2 {\rm Sc}}\biggr] ,
 \label{C6}
\end{eqnarray}
where $\sigma_{\rm v}$ is given by
Eq.~(\ref{R8}), $\sigma_{_{T}} \approx 1$ and
$m=1, 2, 3, ...$. The first mode $(m=1)$ has the
minimum threshold for the excitation of the
tangling clustering instability.

The tangling clustering instability depends on
the ratio $\sigma_{_{T}} / \sigma_{\rm v}$. For
the $ \delta $-correlated in time random Gaussian
compressible velocity field
$\sigma_{_{T}}=\sigma_{\rm v}$ (for details, see
Refs.~\onlinecite{EKR02}). In this case the
second moment $\Phi(t,{\bm R})$ can only decay,
in spite of the compressibility of the velocity
field. On the contrary, for the finite
correlation time of the turbulent velocity field
$\sigma_{_{T}} \not=\sigma_{\rm v}$, and  the
correlation function $\Phi(R)$ grows
exponentially in time, i.e., the tangling
clustering instability is excited.

\begin{figure}
\vspace*{1mm} \centering
\includegraphics[width=9cm]{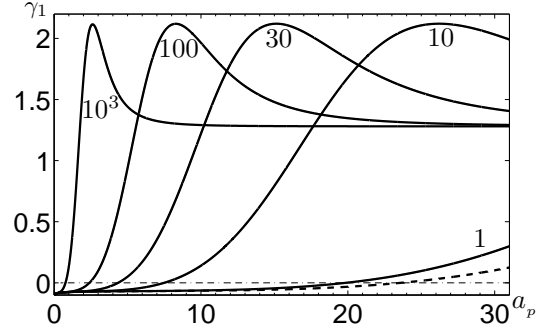}
\caption{\label{Fig1} The growth rate $\gamma_1$
of the tangling clustering instability (in units
of $1 / \tau_\eta$) of the first mode $(m=1)$
versus the particle radius $a_p$ for different
values of parameter $\Gamma$, and
$\sigma_{_{T}}=1$, ${\rm Sc}=10^6 a_p$. The
particle radius $a_p$ is given in $\mu$m. The
dashed line corresponds to the inertial
clustering instability $(\Gamma=1)$.}
\end{figure}

\begin{figure}
\vspace*{1mm} \centering
\includegraphics[width=9cm]{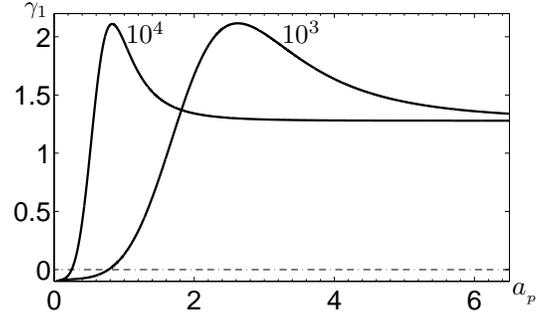}
\caption{\label{Fig2} The growth rate $\gamma_1$
of the tangling clustering instability of the
first mode $(m=1)$ versus the particle radius
$a_p$ for larger values of parameter $\Gamma$,
and $\sigma_{_{T}}=1$, ${\rm Sc}=10^6 a_p$. The
particle radius $a_p$ is given in $\mu$m.}
\end{figure}

Figures~\ref{Fig1}--\ref{Fig2} show the growth
rate [see Eq.~(\ref{C6})] of the tangling
clustering instability of the first mode $(m=1)$
versus the particle radius $a_p$ for different
values of parameter $\Gamma$. One can see from
these figures that the characteristic time of the
tangling clustering instability is of the order
of the Kolmogorov time scale (the growth rate
$\gamma_1$ in Figs.~\ref{Fig1}--\ref{Fig2} is
measured in units of the inverse  Kolmogorov
time). Remarkably, for every parameter
$\Gamma({\rm Re}, {\rm Ma})$ there is a rather
sharp maximum of the function $\gamma_1(a_p)$.
This implies that depending on the parameters of
the turbulence there is a preferential particle
size for which the particle clustering due to the
excitation of the tangling clustering instability
is much faster than for other values of the
particle size. Moreover, the growth rate of the
tangling clustering instability is much larger
than that of the inertial clustering instability
(the growth rate for the inertial clustering
instability in turbulence with a zero mean
temperature gradient is shown in Fig.~\ref{Fig1}
by the dashed line).

\section{Solution for tangling clustering instability with non-zero source
term $(I \not= 0)$}

In this Section we obtain solution of
Eq.~(\ref{T1}) which includes both, the tangling
clustering instability and the source term for
the tangling clustering. This implies that we
consider solution of Eq.~(\ref{T1}) in the
vicinity of the thresholds of the excitation of
the tangling clustering instability.

Let us consider the {\it turbulent diffusion
range of scales}, $1 / \sqrt{\rm Sc} \ll R \ll 1$
and introduce the following function $\Psi(t,
R)=\Phi(t, R) \, R^{1 + \mu}$, where
\begin{eqnarray}
\mu={1 + \sigma_{_{T}} \over 1 +3
\sigma_{_{T}}}\left(1 -2 \sigma_{_{T}} +
{10\sigma_{\rm v} \over 1+\sigma_{\rm v}}
\right).
 \label{E1}
\end{eqnarray}
Equation~(\ref{T1}) is reduced to the Schrodinger
type equation:
\begin{eqnarray}
{\partial \Psi(t,R) \over \partial t} = {1 \over
M(R)} \Psi'' - {\cal U} \, \Psi + I(R) ,
 \label{E2}
\end{eqnarray}
with the potential ${\cal U}$ and $1 / M(R)$ in
the form:
\begin{eqnarray}
&&{\cal U} = -{2 \over 3}\biggl[{40\sigma_{\rm v}
\over 1+\sigma_{\rm v}}\left(1 +{5\sigma_{\rm v}
\over (1+\sigma_{\rm v})} {(1 + \sigma_{_{T}})
\over (1 +3 \sigma_{_{T}})}\right)
\nonumber\\
&& \quad +(2 \sigma_{_{T}}-1) \left(1 - (2
\sigma_{_{T}}-1){1 + \sigma_{_{T}} \over 1 +3
\sigma_{_{T}}}\right) \biggr],
 \nonumber\\
 \label{E3}\\
&&{1 \over M(R)} = {2 \over 3} {(1 +
3\sigma_{_{T}}) \over (1 + \sigma_{_{T}})} \,
 R^2 ,
 \quad \quad
I(R) =B \, R^{1 + \mu} ,
 \label{E4}
\end{eqnarray}
and $B=20\sigma_{\rm v} / (1+\sigma_{\rm v})$.
Here we took into account that the main
contribution to the source term $I(R)$ for large
Reynolds numbers is due to the first term $B({\bm
R}) N^2$ in Eq.~(\ref{B9}). Other contributions
are negligible $\sim 10 {\rm Re}^{-3/2} \ln^2{\rm
Re}$ [see Eqs.~(\ref{D9})
and~(\ref{R8})-(\ref{R9})].

We choose the initial conditions which correspond
to a turbulence without particle clusters:
$\Psi(t=0,R)=0$. Now we seek a solution of
Eq.~(\ref{E2}) in the following form:
$\Psi(t,R)=\sum_{m=1}^{\infty} f_m(t) \Psi_m(R)$
and $I(R)=B \sum_{m=1}^{\infty} A_m \Psi_m(R)$,
where $\Psi_m(R)$ are the eigenfunctions
determined by the following equation with
$I(R)=0$:
\begin{eqnarray}
{1 \over M(R)} \Psi_m'' - ({\cal U}+\gamma_m) \,
\Psi_m =0.
 \label{E11}
\end{eqnarray}
Consequently, the function $f_m(t)$ is determined
by the following equation:
\begin{eqnarray}
{\partial f_m \over \partial t} = \gamma_m \, f_m
+ B\, A_m .
 \label{E12}
\end{eqnarray}
Taking into account the orthogonality of the
eigenfunctions,
\begin{eqnarray}
{\int_{0}^{\infty} M(R) \Psi_m(R) \Psi_n(R) \, dR
\over \int_{0}^{\infty} M(R) \Psi_m^2(R) \, dR} =
\delta_{mn},
 \label{E10}
\end{eqnarray}
and solving Eq.~(\ref{E12}) we obtain the
expressions for the function $f_m(t)$ and $A_m$:
\begin{eqnarray}
&& f_m(t) = {20 \sigma_{\rm v} \, A_m \over
(1+\sigma_{\rm v})  \, \gamma_m} \,
\left[\exp(\gamma_m t) - 1 \right] ,
 \label{E5}\\
&& A_m = {\int_{0}^{\infty} M(R) R^{1 + \mu}
\Psi_m(R) \, dR \over \int_{0}^{\infty} M(R)
\Psi_m^2(R) \, dR}.
 \label{E6}
\end{eqnarray}
Taking into account that
$\Phi_m(R=a_p)=\left[\Psi_m(t, R) \, R^{-(1 +
\mu)} \right]_{R=a_p}=1$ we obtain the
correlation function $\Phi(R=a_p,t)$ in the form:
\begin{eqnarray}
&& \Phi(R=a_p,t) = {20 \sigma_{\rm v} \over
1+\sigma_{\rm v}} \, \sum_{m=1}^{\infty} {A_m
\over \gamma_m} \left[\exp(\gamma_m t) - 1
\right] .
 \label{E14}
\end{eqnarray}
When the tangling clustering instability is
excited, it causes formation of a cluster with
the particle number density inside the cluster,
which is much larger than the mean particle
number density.

The solution for $\Phi(t,R)$ [see
Eq.~(\ref{C31})], which satisfies the above
conditions has the following dimensional form:
\begin{eqnarray}
\Phi(t,R) &=& N^2 \, {20 \sigma_{\rm v} \over
1+\sigma_{\rm v}} \, \sum_{m=1}^{\infty} {A_m
\over \gamma_m} \left({R \over
\ell_D}\right)^{-\lambda}
\nonumber\\
&& \times \cos\left[ \kappa \ln \left({R \over
\ell_D}\right) \right] \, \left[\exp(\gamma_m t)
- 1 \right] .
 \label{E8}
\end{eqnarray}
Here the correlation function $\Phi(t,R)$ has the
global maximum at $R=\ell_D/\ell_\eta$, i.e., we
assumed that in the molecular diffusion region
the correlation function $\Phi(t,R)$ is nearly
constant, $\Phi(t,R) \approx 1$ [see
Eq.~(\ref{C3})]. The first minimum of the
correlation function $\Phi(t,R)$ for the mode
$m=1$ is located in $R=R_{\rm
min}=\exp(1/\lambda)$, and it is given by the
following expression
\begin{eqnarray}
{\Phi_{\rm min} \over N^2} = - {B \, A_1 \kappa
\over e \, \gamma_1 \, \lambda} \, {\rm
Sc}^{-\lambda/2} \, \left[\exp(\gamma_1 t) - 1
\right] ,
 \label{E16}
\end{eqnarray}
where we took into account that $\cos\left[
\kappa \ln (R_{\rm min} / \ell_D)\right] = -
\kappa / \lambda$. On the other hand, the maximum
value of the correlation function $\Phi(t,R)$ is
\begin{eqnarray}
{\Phi_{\rm max} \over N^2} = {B \, A_1 \over
\gamma_1} \, \left[\exp(\gamma_1 t) - 1 \right] ,
 \label{E17}
\end{eqnarray}
where $\Phi_{\rm max}=\Phi(R=a_p,t)$ [see
Eq.~(\ref{E14})]. Equations~(\ref{E16}) and
(\ref{E17}) yield:
\begin{eqnarray}
{\Phi_{\rm min} \over \Phi_{\rm max}} = - {\pi
\over e \, \lambda} \left({{\rm Sc}^{-\lambda/2}
\over \ln {\rm Sc}}\right) .
 \label{E9}
\end{eqnarray}
Since $n_p=N+n\geq 0$, the function $\Phi(t,{\bm
R})\equiv \langle n(t,{\bm x}) n(t,{\bm y})
\rangle = \langle n_p(t,{\bm x}) n_p(t,{\bm y})
\rangle - N^2 \geq -N^2$. Therefore, the minimal
possible value of the function $\Phi(t,{\bm R})$
is $\Phi_{\rm min} =-N^2$. This condition
together with Eq.~(\ref{E9}) allow us to estimate
the maximum number density of particles attained
inside the cluster during the tangling clustering
instability:
\begin{eqnarray}
{n_p^{\rm max} \over N} = \left(1 + {e \, \lambda
\over \pi} \, {\rm Sc}^{\lambda/2} \, \ln {\rm
Sc}\right)^{1/2} .
 \label{E15}
\end{eqnarray}

\begin{figure}
\vspace*{1mm} \centering
\includegraphics[width=9cm]{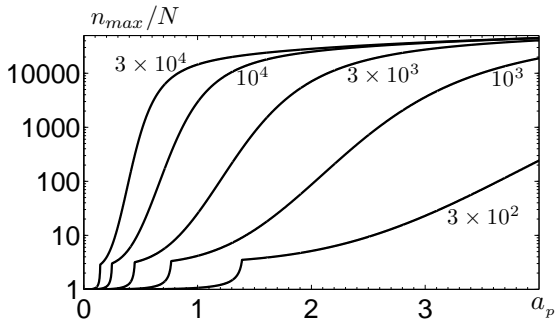}
\caption{\label{Fig3} The particle number density
inside the cluster $n_{\rm max} / N$ versus the
particle radius $a_p$ for different values of
parameter $\Gamma$, and $\sigma_{_{T}}=1$, ${\rm
Sc}=10^6 a_p$. The particle radius $a_p$ is given
in $\mu$m.}
\end{figure}

The maximum value of the particle number density
inside the cluster, $n_{\rm max} / N$, versus the
particle radius is shown in Fig.~\ref{Fig3} for
different values of parameter $\Gamma$. The
discontinuity of the first derivative of $n_{\rm
max} / N$ which is seen in Fig.~\ref{Fig3} is
related to the transition from one mechanism of
particle tangling clustering due to the source
term to another mechanism caused by the tangling
clustering instability. The exponential growth at
the linear stage of the instability is saturated
by the nonlinear effects. The values of $n_{\rm
max} / N$ in Fig.~\ref{Fig3} are calculated using
Eq.~(\ref{E15}) that takes into account possible
saturation of the tangling clustering instability
caused by the exhaustion of the particles in the
region surrounding the cluster. Inspection of
Fig.~\ref{Fig3} shows that the particle number
density inside the cluster can increase by a
factor of $10^4$ in comparison with the mean
particle number density.

There are also other mechanisms leading to the
nonlinear saturation of the tangling clustering
instability discussed in detail in
Ref.~\onlinecite{EKR02}. However, as follows from
our analysis the main significant mechanism of
saturation of the growth of the tangling
clustering instability for small particles is
exhaustion of the particles in the surrounding
area. Indeed, the tangling clustering instability
causes strong redistribution of particles so that
inside the clusters the particle number density
strongly increases at some instant, while in the
surrounding regions it decreases. With the
decrease of the number density of particles the
hydrodynamic description becomes inapplicable. It
should be noted that we consider situation when
there is only the redistribution of the particles
without their creation or annihilation. This
implies that particles from the cluster vicinity
are concentrated in the central part of it, which
can be expressed using the Corrsin integral of
the correlation function of the particle number
density fluctuations: $\int_{0}^{\infty}
\Phi(t,R) R^{2} \, dR =0$. This condition implies
that the tail of the correlation function
$\Phi(t,R)$ must be negative [i.e., there is the
anti-correlation tail of the function
$\Phi(t,R)$]. The transition from central
positive part $\Phi(t,R)>0$ to the negative tail
of $\Phi(t,R)$ occurs at the distance that is of
the order of several Kolmogorov scales. Note,
that in contrast to the inertial clustering, the
tangling clustering instability accumulates
particles into the cluster from the scales which
are much larger than the Kolmogorov length scale.
The reason is that the tangling mechanism
generates fluctuations of the particle number
density in all scales of inertial range in
turbulence with imposed mean temperature
gradient. Consequently, the concentration of
particles inside the cluster increases due to the
tangling clustering instability by several orders
of magnitude.

\begin{figure}
\vspace*{1mm} \centering
\includegraphics[width=9cm]{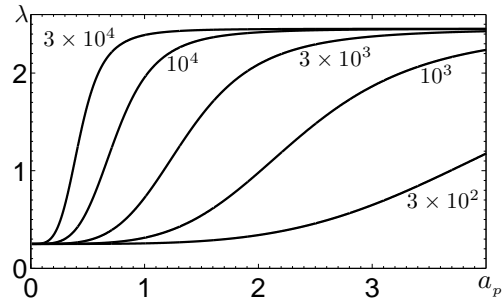}
\caption{\label{Fig4} The exponent $\lambda$
versus the particle radius $a_p$ for different
values of parameter $\Gamma$ and
$\sigma_{_{T}}=1$. The particle radius $a_p$ is
given in $\mu$m.}
\end{figure}

The value $n_{\rm max}$ depends strongly on the
Schmidt number ${\rm Sc}$  and on the exponent
$\lambda$ [see Eq.~(\ref{E15})]. The exponent
$\lambda$ depends on the degree of
compressibility of the particle velocity field,
$\sigma_{\rm v} \propto {\rm St}^2 \, \Gamma({\rm
Ma}, {\rm Re}, \ell_0/L_T)$ [see Eq.~(\ref{R9})].
The calculated values of exponent $\lambda$
versus the particle radius for different values
of parameter $\Gamma$ are shown in
Fig.~\ref{Fig4}, that explains strong dependence
of the maximum particle number density inside the
cluster $n_{\rm max}$ on the particle radius and
the parameter $\Gamma$.

In general, other nonlinear mechanisms may limit
the growth of the tangling clustering instability
in the nonlinear stage of its evolution, and
therefore they limit the maximum attainable value
of the particle number density, $n_p^{\rm max}$,
inside the cluster. For example, a momentum
coupling of particles and turbulent fluid becomes
essential when the mass loading parameter
$\varsigma= m_{\rm p} n_{\rm max}/\rho$ is of the
order of unity. \cite{CST11} Introducing a mean
particle mass density $\overline{\rho}_p = m_p \,
N$ (e.g., in cloud physics it is a liquid water
content measured in g/cm$^3$), we obtain the
following constraint: $n_p^{\rm max} / N \leq
\varsigma \rho / \overline{\rho}_p$. Using $\rho
=1.3 \times 10^{-3}$ g/cm$^3$ and
$\overline{\rho}_p = 10^{-6}$ g/cm$^3$ we arrive
at $n_p^{\rm max} / N \leq 1300$ for $\varsigma
=1$. This limiting effect should be taken into
account together with the saturation mechanism
caused by the exhaustion of the particles in the
vicinity of the cluster.

\begin{figure}
\vspace*{1mm} \centering
\includegraphics[width=9cm]{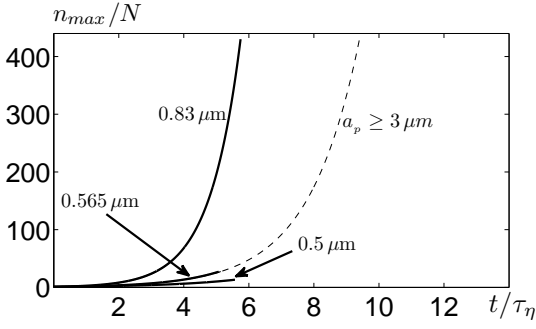}
\caption{\label{Fig5} The particle number density
inside the cluster $n_{\rm max} / N$ as a
function of time for different values of the
particle radius $a_p$ and $\Gamma=10^4$,
$\sigma_{_{T}}=1$, ${\rm Sc}=10^6 a_p$. The
particle radius $a_p$ is given in $\mu$m. The
dashed line is for $a_p \geq 3 \mu$m.}
\end{figure}

Figure~\ref{Fig5} shows the temporal evolution of
the number density of particles inside the
cluster during the excitation of the tangling
clustering instability for particles of different
radii. Figure~\ref{Fig5} reveals several
interesting features pertinent to the instability
which deserve to be mentioned. The tangling
clustering instability is less effective for very
small particles, $a_p \leq 0.5\mu$m. The reason
is that this instability is saturated by the
exhaustion of the particles in the vicinity of
the cluster for very low value of $n_{\rm max}/N
\approx 13.2$. For the particles having
sub-micron and micron sizes the concentration
inside the cluster can increase due to the
tangling clustering instability by several orders
of magnitude. On the other hand, as follows from
Fig.~\ref{Fig2} the growth rate of the tangling
clustering instability is the same for particles
with $a_p=0.565 \mu$m and for all particles with
$a_p \geq 3 \mu$m. However, contrary to the case
$a_p \geq 3 \mu$m, the saturated value of the
particle number density enhancement due to the
instability for $a_p=0.565 \mu$m is very low,
$n_{\rm max}/N \approx 27.2$.

\section{Effect of droplet evaporation on tangling clustering
instability}

Let us study the effect of droplet evaporation on
tangling clustering instability in stably
stratified turbulence. The equation for the
instantaneous number density $n_p(t,{\bf x})$  of
droplets of the radius $a_p$ reads:
\begin{eqnarray}
\frac{\partial n_p}{\partial t} &+& {\bf \nabla
\cdot} (n_p \,{\bf v}) = D_m\, \Delta n_p - {n_p
\over \tau_{\rm ev}} + I_0 ,
\nonumber\\
 \label{AP1}
\end{eqnarray}
where the second term in the right hand part of
Eq.~(\ref{AP1}) takes into account the droplet
evaporation with the characteristic time
$\tau_{\rm ev}$, and the last term,  $I_0$,
describes source of droplets due to condensation,
which for simplicity is assumed to be constant.
The equation for fluctuations of the droplet
number density, $n(t,{\bf x})= n_p(t,{\bf x}) -
N(t,{\bf x})$, reads:
\begin{eqnarray}
&&\frac{\partial n}{\partial t} + {\bf \nabla
\cdot} (n \,{\bf v}- \langle n \,{\bf v}\rangle)
= D_m\, \Delta n - ({\bf v} {\bf \cdot \nabla}) N
- N \, {\bf \nabla \cdot} \,{\bf v}
\nonumber\\
&& - {n \over \tau_{\rm ev}} .
 \label{AP2}
\end{eqnarray}
The last term in Eq.~(\ref{AP2}) describes the
droplet evaporation. Using Eq.~(\ref{AP2}) we
derive equation for the evolution of the
two-point second-order correlation function of
the droplet number density, $\Phi(t,{\bf R})$,
see Eq.~(\ref{B2}), in which
\begin{eqnarray}
D_{ij}^{^{T}}({\bf R}) &\approx& 2
\int_{0}^{\infty} \langle v_{i}
\big[0,\bec{\xi}(t,{\bf x}|0)\big] \,
v_{j}\big[\tau,\bec{\xi}(t,{\bf x}+{\bf
R}|\tau)\big] \rangle
\nonumber\\
&& \times G(\tau)  \,d \tau ,
 \label{AP4}\\
B({\bf R}) &\approx& 2 \int_{0}^{\infty} \langle
b\big[0,\bec{\xi}(t,{\bf x}|0)\big]
\,b\big[\tau,\bec{\xi}(t,{\bf x}+{\bf
R}|\tau)\big] \rangle
\nonumber\\
&& \times G(\tau)  \,d \tau ,
 \label{AP5}\\
U_{i}({\bf R}) &\approx& -2 \int_{0}^{\infty}
\langle v_{i}\big[0,\bec{\xi}(t,{\bf x}|0) \big]
\,b\big[\tau,\bec{\xi}(t,{\bf x}+{\bf
R}|\tau)\big] \rangle
\nonumber\\
&& \times G(\tau)  \,d \tau ,
 \label{AP6}
\end{eqnarray}
$G(\tau)=\exp(-\tau/\tau_{\rm ev})$, and other
terms do not change in case of the droplet
evaporation.

Equation for $\Phi(t,{\bf R})$ can be rewritten
in the dimensionless form as follows:
\begin{eqnarray}
{\partial \Phi\over \partial t} &=& {1 \over
M(R)} \left[\Phi'' + 2 \, \left({1 \over R} +
\chi(R) \right) \, \Phi' \right]
\nonumber\\
&& + \left[B(R)-{2 \tau_{_{\rm D}} \over
\tau_{\rm ev}} \right]\, \Phi + I(R) ,
 \label{AP7}
\end{eqnarray}
where distance $R$ is measured in units of
Kolmogorov scale $\ell_\eta$ and time $t$ is
measured in units of $\tau_\eta^2/\tau^{\rm
eff}\equiv\tau_{_{\rm D}}/3$. Here the effective
time $\tau^{\rm eff}$ is determined by the
following expression:
\begin{eqnarray}
\tau^{\rm eff}=  {\tau_\eta \tau_{\rm ev} \over
\tau_\eta+\tau_{\rm ev}},
 \label{AP8}
\end{eqnarray}
and a modified turbulent diffusion time is
determined as $\tau_{_{\rm
D}}=\ell_\eta^2/D_{_{\rm T}}^{\rm eff}$, the
effective turbulent diffusion coefficient
$D_{_{\rm T}}^{\rm eff}$ in the Kolmogorov scale
reads \cite{EKR98}
\begin{eqnarray}
D_{_{\rm T}}^{\rm eff}= {\tau^{\rm eff} u_\eta^2
\over 3} ,
 \label{AP9}
\end{eqnarray}
and a modified function $1/M(R)$ is
\begin{eqnarray}
{1 \over M(R)} = {2 \over {\rm Sc}^{\rm eff}} +
{2 \over 3} [1 - F - (R F_c)'],
 \label{AP10}
\end{eqnarray}
where ${\rm Sc}^{\rm eff}={\rm Sc} \, \tau^{\rm
eff} / \tau_\eta$. Equation~(\ref{AP7}) shows
that the evaporation decreases the term $[B(R)-2
\tau_{_{\rm D}} / \tau_{\rm ev}]$ which is
responsible for the generation of fluctuations of
the droplet number density. Equation~(\ref{AP9})
has a simple physical meaning. In the case when
the droplet evaporation time is much smaller than
the turbulent correlation time, the turbulent
diffusion coefficient is renormalized as given by
Eq.~(\ref{AP9}), for details see
Ref.~\onlinecite{EKR98}.

\begin{figure}
\vspace*{1mm} \centering
\includegraphics[width=9cm]{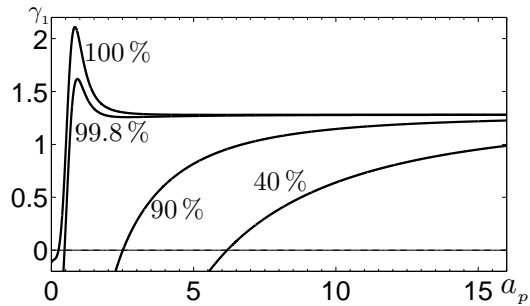}
\caption{\label{Fig6} The growth rate $\gamma_1$
of the tangling clustering instability (in units
of $1 / t_\eta$) of the first mode $(m=1)$ versus
the particle radius $a_p$ for different values of
the relative humidity $\phi$, and $\Gamma=10^4$,
$\sigma_{_{T}}=1$, ${\rm Sc}=10^6 a_p$. The
particle radius $a_p$ is given in $\mu$m. The
dashed line corresponds to the inertial
clustering instability.}
\end{figure}

The analysis similar to that performed in Sect.~4
yields the growth rate for the mode $m$ of the
tangling clustering instability:
\begin{eqnarray}
\gamma_m &=& {1 \over 3 (1 + 3 \sigma_{_{T}})}
\biggl[{200 \sigma_{\rm v} (\sigma_{_{T}} -
\sigma_{\rm v}) \over (1 + \sigma_{\rm v})^2} -
{(3 - \sigma_{_{T}})^{2} \over 2 (1 +
\sigma_{_{T}})}
\nonumber\\
&& - {2 \pi^2 m^2 (1 + 3 \sigma_{_{T}})^{2} \over
(1 + \sigma_{_{T}}) \ln^2 {\rm Sc}^{\rm
eff}}\biggr] - {2\tau_\eta^2 \over \tau^{\rm eff}
\tau^{\rm ev}} ,
 \label{AP11}
\end{eqnarray}
where $m=1, 2, 3, ...$. This growth rate
$\gamma_m$ of the second moment of particles
number density was obtained by matching the
correlation function $\Phi(R)$ and its first
derivative $\Phi'(R)$ at the points $R = 1 /
\sqrt{{\rm Sc}^{\rm eff}}$ and $R = 1$, which
also yields: $\kappa / 2(C_{1} + C _{2}) \approx
\pi \, m / \ln {\rm Sc}^{\rm eff}$. Therefore,
the evaporation of droplets causes decrease of
the instability growth rate. Figure~\ref{Fig6}
shows the growth rate $\gamma_1$ of the tangling
clustering instability for different values of
the relative humidity $\phi$ versus the particle
radius. Here we used the following expression for
the evaporation time of droplets $\tau^{\rm ev}=
2.1 \times 10^{-3} a_p^2 / (1-\phi)$, where the
droplet radius is in microns and time is in
seconds (see, e.g., Ref.~\onlinecite{SP06}).
Inspection of Fig.~\ref{Fig6} shows that the
evaporation of droplets strongly affects the
tangling clustering instability for small
droplets, i.e., it increases the instability
threshold in the droplet radius depending on the
relative humidity $\phi$. In addition, there is
sharp maximum of the growth rate for 1 $\mu$m
droplets if the relative humidity is close to the
supersaturated values: 99.8 \% and 100 \%. For
low relative humidity the growth rate of the
tangling clustering instability is less in
comparison with the supersaturated case.

\begin{figure}
\vspace*{1mm} \centering
\includegraphics[width=9cm]{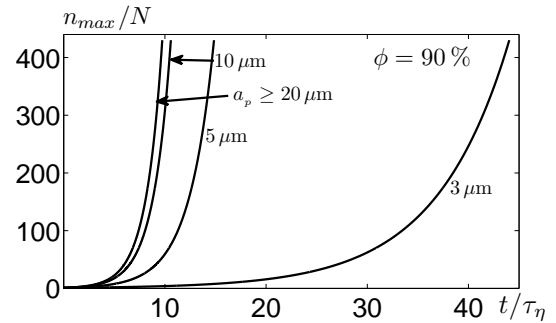}
\caption{\label{Fig7} The particle number density
inside the cluster $n_{\rm max} / N$ as a
function of time for different values of the
particle radius $a_p$, and the relative humidity
$\phi=90 \%$, $\Gamma=10^4$, $\sigma_{_{T}}=1$,
${\rm Sc}=10^6 a_p$. The particle radius $a_p$ is
given in $\mu$m.}
\end{figure}

Figure~\ref{Fig7} shows the temporal evolution of
the number density of particles inside the
cluster during the excitation of the tangling
clustering instability for the relative humidity
$\phi=90 \%$ and different droplet radius. This
figure also demonstrates that the evaporation of
droplets strongly affects the tangling clustering
instability.

\section{Discussion and Conclusions}

The present study has been inspired by the
previous work \cite{EKR10}, where it was shown
that the tangling clustering of inertial
particles in the temperature stratified
turbulence holds the potential to promote a
strong clustering with the considerably enhanced
particle concentration inside the cluster. In
this study based on the thorough theoretical
analysis, it is demonstrated that the temperature
fluctuations strongly contribute to the tangling
clustering instability. Temperature fluctuations
caused by tangling of the mean temperature
gradient by the velocity fluctuations, produce
pressure fluctuations and enhance considerably
particle clustering. The growth rate of the
tangling clustering instability is by a factor of
$\sqrt{\rm Re} \, (\ell_0 / L_T)^2 / (3 {\rm
Ma})^4$ larger than the growth rate of the
inertial clustering instability.

The growth rate of the tangling clustering
instability and the particle number density
inside the cluster after saturation of the
instability on the nonlinear stage depends on the
parameter $\Gamma \approx \left({\rm Re}^{1/4} /
9 \, {\rm Ma}^{2} \right) \,\ell_0 \left|{\bm
\nabla} T\right| / T$. We also found that
depending on the parameters of turbulence and the
mean temperature gradient there is the
preferential clustering of particles of a
particular size (the growth rate of the tangling
clustering instability has a sharp maximum at
this size). The growth of the particle number
density inside the cluster caused by the tangling
clustering instability is significantly larger
(by several orders of magnitudes) than the
increase of the particle number density inside
the cluster caused by the source tangling
clustering.\cite{EKR10}

We demonstrated the strong effect of the droplet
evaporation on this instability. The tangling
clustering instability in the temperature
stratified turbulence may enhance significantly
the collision rate of small particles, which is
of interest for atmospheric physics and many
other practical applications. In particular this
effect can substantially accelerate the
coalescence of small droplets in atmospheric
turbulence with temperature gradients.

\begin{acknowledgements}
This research was supported in part by the Israel
Science Foundation governed by the Israeli
Academy of Sciences (Grant 259/07), by EU COST
Actions MP0806 and ES1004, by the EC FP7 project
ERC PBL-PMES (Grant 227915), by the Russian
Government Mega Grant (Grant 11.G34.31.0048), and
by the Grant of Russian Ministry of Science and
Education (Contract No. 8648).
\end{acknowledgements}

\end{document}